\documentclass[preprint,5p,times,twocolumn]{elsarticle}

\usepackage{amssymb}
\usepackage{graphicx}
\usepackage{amsmath,amsfonts,amssymb}
\usepackage{citesort}
\usepackage{epsfig}
\usepackage[hyperindex,breaklinks]{hyperref} 
\usepackage{latexsym}
\usepackage{graphicx}
\usepackage{amsmath}
\usepackage{amsfonts}   
\usepackage{amssymb}    
\usepackage{float}
\usepackage{bm}
\usepackage{hyperref}
\usepackage{color}
\usepackage[columnwise]{lineno}

\definecolor{purple}{rgb}{0.5,0,0.5}
\definecolor{dg}{rgb}{0.0, 0.5, 0.0}

\def\lsim{\raise0.3ex\hbox{$\;<$\kern-0.75em\raise-1.1ex\hbox{$\sim\;$}}}
\def\gsim{\raise0.3ex\hbox{$\;>$\kern-0.75em\raise-1.1ex\hbox{$\sim\;$}}}
\def    \beq            {\begin{equation}}
\def    \eeq            {\end{equation}}
\def    \bea           {\begin{eqnarray}}
\def    \eea           {\end{eqnarray}}

\def\minv{m^{inv}_{\gamma\gamma}}
\def\fbi{\rm fb^{-1}}
\def\ecm{\rm E_{\rm CM}}
\def\hx{{H}_{\rm X}}

\def\ecm{\rm E_{\rm CM}}

\journal{Physics Letters B}

\begin{document}

\begin{frontmatter}

\title{Di-photon resonance around 750 GeV: shedding light on the theory underneath\tnoteref{label1}}
\tnotetext[label1]{LPT-Orsay-15-101, CPHT-RR060.1215, HRI-RECAPP-2015-020}

\author[jd1]{Joydeep Chakrabortty}
\ead{joydeep@iitk.ac.in} 
\address[jd1]{Department of Physics, Indian Institute of Technology, Kanpur-208016, India}

\author[ac1,ac2]{Arghya Choudhury}
\ead{a.choudhury@sheffield.ac.uk} 
\address[ac1]{Consortium for Fundamental Physics, Department of Physics and Astronomy, \\
University of Sheffield, Sheffield S3 7RH, United Kingdom}
\address[ac2]{Consortium for Fundamental Physics, Department of Physics and Astronomy, \\
University of Manchester, Manchester, M13 9PL, United Kingdom}

\author[pg1,pg2]{Pradipta~Ghosh}
\ead{pradipta.ghosh@th.u-psud.fr} 
\address[pg1]{Laboratoire de Physique Th\'eorique, CNRS$^1$\fnref{label3}, 
Univ.  Paris-Sud, Universit\'e Paris-Saclay, 91405 Orsay, France}
\fntext[label3]{UMR 8627}
\address[pg2]{Centre de Physique Th\'eorique, \'Ecole polytechnique, CNRS$^2$\fnref{label4}, 
Universit\'e Paris-Saclay, 91128 Palaiseau, France}
\fntext[label4]{UMR 7644}

\author[sm1]{Subhadeep Mondal}
\ead{subhadeepmondal@hri.res.in}
\address[sm1]{Regional Centre for Accelerator-based Particle Physics,
Harish-Chandra Research Institute, Allahabad 211019, India}

\author[jd1]{Tripurari Srivastava}
\ead{tripurar@iitk.ac.in}

\begin{abstract}
Both the ATLAS and CMS collaborations have recently observed an excess
in the di-photon invariant mass distribution in the vicinity of $750$ GeV 
with a local significance of $\sim3\sigma$. In this article we try to investigate this excess 
in the context of a minimal simplified framework assuming effective interactions
of the hinted resonance with photons and gluons. We scrutinise
the consistency of this observation with possible accompanying 
yet hitherto unseen signatures of this resonance. Subsequently, we
try to probe the nature of new particles, e.g., spin, electric charge and
number of colour, etc., that could remain instrumental to explain
this excess through loop-mediation. 
\end{abstract}



\end{frontmatter}


The recent observation by the LHC collaborations \cite{ATLAS-run-II-1,CMS-run-II-2,
ATLAS-run-II-1L,CMS-run-II-2L}, concerning an excess in the di-photon invariant mass 
distribution $\minv$ near $750$ GeV, has gained huge
attention in the particle physics community. The ATLAS group, 
using $3.2$ $\fbi$ of data with $13$ TeV centre-of-mass energy $(\ecm)$, 
has estimated a local (global) significance of $3.9\sigma~(2.0\sigma)$ for a 
mass of the resonance $M_X=750$ GeV \cite{ATLAS-run-II-1L}. At the same time,
the CMS collaboration has noticed
a local (global) significance of $2.8\sigma-2.9\sigma~(<1.0\sigma)$ 
for $M_X=760$ GeV \cite{CMS-run-II-2L}
using $3.3$ $\fbi$ of data at $\ecm=13$ TeV. Combining with the 
run-I data ($19.7\,\fbi$ at $\ecm=8$ TeV),
the CMS excess appears at $M_X=750$ GeV \cite{CMS-run-II-2L} with a 
local (global) significance of $3.4\sigma~(1.6\sigma)$. The latter
corresponds to a narrow width for the resonance, $\Gamma_X=105$ MeV
while interpretation with only $13$ TeV data indicates $\Gamma_X=10.6$ GeV.
The ATLAS measurement, on the contrary, hints a large decay
width $\Gamma_X=45$ GeV \cite{ATLAS-run-II-1L}.

This is the first surprise from LHC run-II with 13 TeV center-of-mass 
energy\footnote{A {\it less significant} increasing fluctuation was also noticed 
during the LHC run-I with 8 TeV center-of-mass 
energy \cite{CMS:2015cwa,Aad:2015mna}.} which remains unexplained
within the Standard Model (SM) framework. In other words,
properties of the said resonance, as experimentally observed so far, e.g.,
excess in $\gamma\gamma$ only and nothing in $ZZ,\,Z\gamma$ or in
di-jet $(jj)$ channels, definitely demand physics beyond the SM (BSM).
It is, thus, timely to explore the origin and associated consequences of this resonance
although the possibility of 
loosing this excess with more data-set can not be completely overlooked. 
A quest to accommodate this excess has already produced a handful of contemporary 
analyses \cite{Harigaya:2015ezk, Mambrini:2015wyu,
Backovic:2015fnp, Angelescu:2015uiz, Nakai:2015ptz, Knapen:2015dap, Buttazzo:2015txu, 
Pilaftsis:2015ycr, Franceschini:2015kwy, DiChiara:2015vdm,Higaki:2015jag, McDermott:2015sck, 
Ellis:2015oso,Low:2015qep,Bellazzini:2015nxw,Gupta:2015zzs,Petersson:2015mkr,Molinaro:2015cwg}
along with a few simultaneous\footnote{Appeared in the arXiv on the same day with this article.} 
\cite{Chao:2015ttq,Cao:2015pto,Kobakhidze:2015ldh,Curtin:2015jcv,Ahmed:2015uqt} studies.
Most of these analyses are proposed within the context of a specific
theory framework, which often requires new decay modes (invisible for example)
and thus, address other issues, for example the dark matter  
(see Refs.~\cite{Mambrini:2015wyu,Backovic:2015fnp}).
We, however, aim to investigate this excess with a simplified effective framework
and will try to explore the nature of hitherto unseen particles
which, while running in the loop, can appear instrumental to produce the observed
di-photon excess.

With this idea we have used a generic Lagrangian which couples
this new resonance $\hx$ with photons and gluons as shown by eq.~(\ref{form1}).
We have further assumed: (1) {\it on-shell} production of 
$H_X$ and (2) a scalar, i.e., spin-0, nature\footnote{The observed
excess is also compatible with a spin-2 nature \cite{CMS-run-II-2,
ATLAS-run-II-1L,CMS-run-II-2L}.} for $H_X$.
The latter is one of the natural options to explain a resonance in di-photon channel, 
i.e., two {\it identical massless spin-1} particles, as 
dictated by Landau-Yang theorem \cite{Landau:1948kw,Yang:1950rg}.
The effective minimal\footnote{We are working in a limit when 
interaction like $\kappa_W W^a_{\mu\nu}W^{\mu\nu}_a$,
i.e., coupling between the SM $SU(2)$ gauge bosons and $\hx$ vanishes.} Lagrangian is written as:

\begin{equation}\label{form1}
{\mathcal{L}}_{ eff} =  \kappa_g G^a_{\mu \nu} G^{\mu \nu}_a {H}_{\rm X}
+ \kappa_A B_{\mu \nu} B^{\mu \nu} {H}_{\rm X},
\end{equation}
%
where $G^a_{\mu\nu}$, $B_{\mu\nu}$ are the associated field strengths
with $``a"$ representing the relevant non-Abelian index. The effective 
$\hx$-$g$-$g$ and $\hx$-$\gamma$-$\gamma$ vertices are parametrised as 
$\kappa_g$ and $\kappa_A$ which encapsulate the effect of new physics appearing in the 
loops. The latter is an absolute necessity since SM-like couplings between
the SM-gauge bosons and $H_X$ appear inadequate \cite{Carena:2012xa} to explain the 
observed {\it{sizable}} decay width $\Gamma_X$ \cite{ATLAS-run-II-1,
ATLAS-run-II-1L} and the production cross-section $\sigma(pp(gg)\to H_X \to \gamma\gamma)$ 
$\sim \mathcal{O}(10$ fb) \cite{ATLAS-run-II-1,CMS-run-II-2,
ATLAS-run-II-1L,CMS-run-II-2L}, consistent with the results of various other LHC searches. 
The observations from different LHC searches put strong constraints on the 
$\kappa_g$ - $\kappa_A$ parameter space. The latter can be translated
in terms of $\hx \to g g$, $\hx \to \gamma \gamma$ branching fractions $(Brs)$ since they are 
$\propto 8\kappa^2_g$, $\kappa^2_A \cos^4\theta_W$, respectively. Moreover, the associated
squared matrix elements are similar while the phase spaces are identical. The number
$`8$' appears from the colour factor and $\theta_W$ is the Weinberg angle \cite{Agashe:2014kda}.

\begin{figure}[t!]
\includegraphics[width=4.39cm,height=3.55cm]{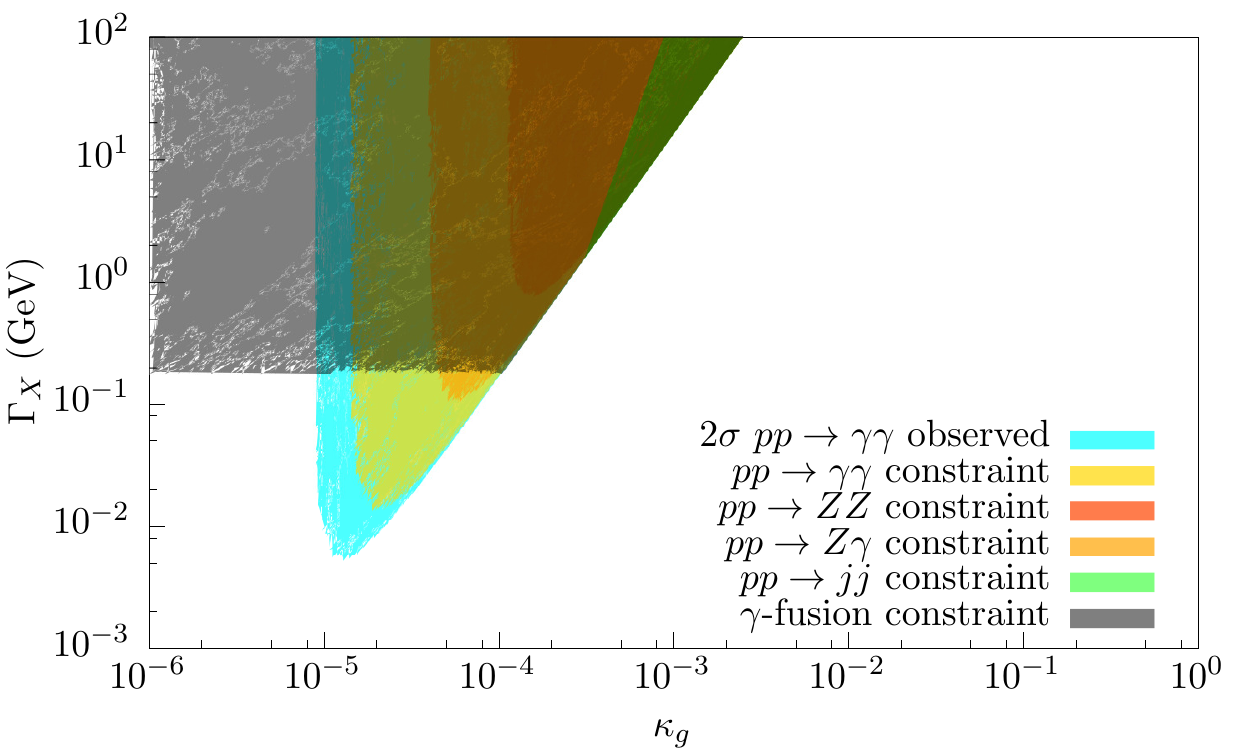}
\includegraphics[width=4.39cm,height=3.55cm]{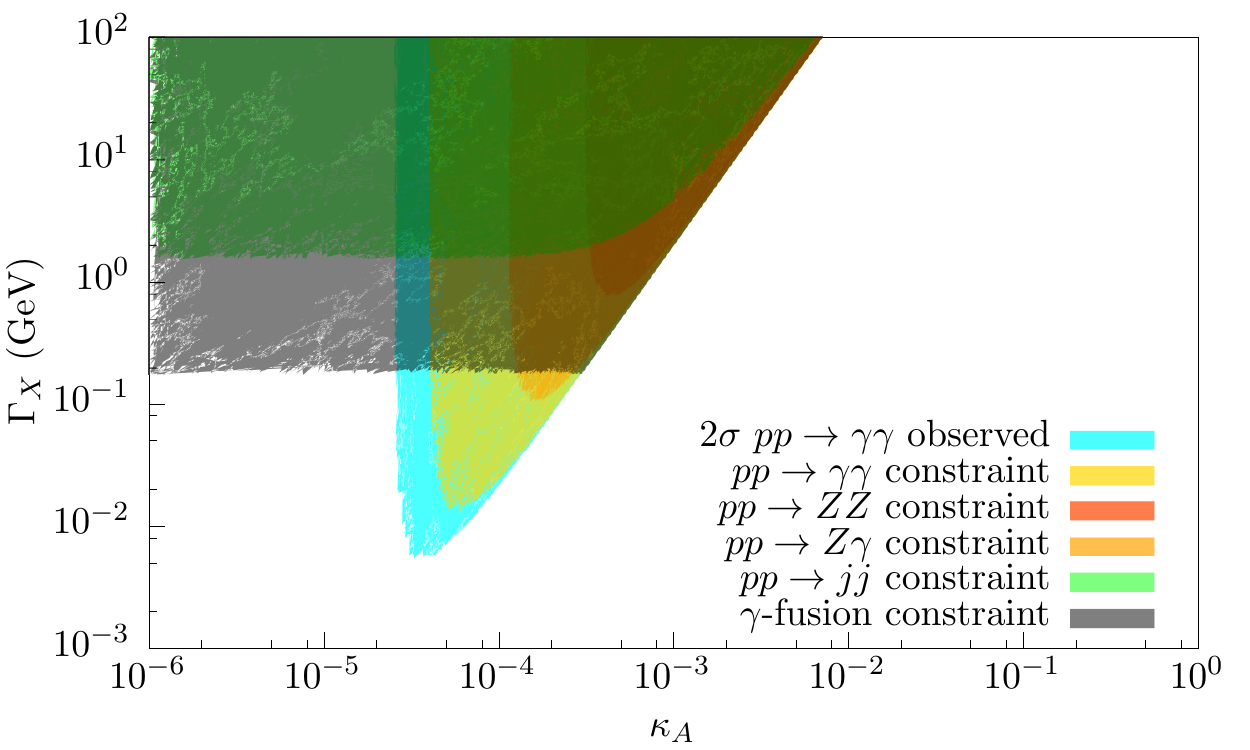}
\caption{Variations of the $\Gamma_X$ (in GeV) with 
$\kappa_g$ (left) and $\kappa_A$ (right).
The details are explained in the text.}
\label{fig:fit1}
\end{figure}

After the electroweak symmetry breaking, the second term of 
eq.~(\ref{form1}) generates effective interactions 
like $\hx\gamma\gamma$ and also $\hx Z \gamma$, $\hx ZZ$,
even with vanishing $\kappa_W$. Their strengths are $\propto$ $\kappa_A\cos^2\theta_W$, 
$\kappa_A\sin\theta_W\cos\theta_W$ and $\kappa_A\sin^2\theta_W$,
respectively. It is thus, important to note that
a non-zero $Br(\hx\to \gamma\gamma)$ would also imply non-zero
$Br(\hx\to Z\gamma,\,ZZ)$ values since all of them are 
connected to $\kappa_A$. Their relative magnitudes, however,
remain different depending on the factor of $\sin\theta_W$
or $\cos\theta_W$. Measurements from the experimental
collaborations for the said processes, using 13 TeV data, remain yet
inadequate\footnote{The ATLAS and CMS collaborations have
recently reported $ZZ$ \cite{atlas-zz13,CMS:2016vvl} and 
$Z\gamma$ \cite{atlas-zph13,CMS:2016rsl} search results with early
13 TeV data.}. Nevertheless, measured information
for $\hx\to ZZ,\, Z\gamma$ and $\hx\to \gamma\gamma$ \cite{ATLAS-diphoton} 
processes from the 8 TeV searches 
definitely constrain the range of $\kappa_A,\,\kappa_g$ parameters.
For example, one obtains $\sigma (pp\to \hx\to ZZ) < 12$ fb \cite{Aad:2015kna}
and $\sigma (pp\to \hx\to Z\gamma) < 11$ fb \cite{Aad:2014fha} from the similar
searches performed by the ATLAS with 8 TeV data. The available parameter 
space is also constrained by the di-jets searches, given as $\sigma(pp\to jj) < 1.9$ 
pb \cite{Aad:2014aqa}\footnote{For di-jet searches, early 13 TeV 
results are also available \cite{ATLAS:2015nsi,Khachatryan:2015dcf}.}, 
such that the missing evidence
of $pp\to\hx\to jj$ process at the 13 TeV appears consistent.
Needless to mention that the CMS collaboration has also
made similar studies \cite{CMS-diphoton,CMS-zz,CMS-zgamma,CMS-dijet}.
Furthermore, if one wishes to account for a large $\Gamma_X$
by introducing new, e.g., invisible decays,
one needs to incorporate the constraints from monojet searches
accordingly \cite{ATLAS-monojet,CMS-monojet}.

In this article we have used the {\it expected} limits 
from 13 TeV LHC searches for $ZZ,\,Z\gamma$,
$jj$ and $\gamma\gamma$ processes, derived using the 8 TeV results. 
We have used {\tt Madgraph~v2.2.3} \cite{Alwall,Alwall:2014hca}
and observed that the production (via gluon fusion)
cross-section with 13 TeV $\ecm$ is roughly
five times of the same with $8$ TeV $\ecm$, i.e., $\sigma(pp\to \hx)|_{13~
\rm TeV}/\sigma(pp\to \hx)|_{8~\rm TeV}\approx 5$, as also noted in Ref.~\cite{Franceschini:2015kwy}.
Further, we have also used the constraint from Ref.~\cite{photon-photon}
assuming that this resonance can also appear through photon fusion.
In our numerical study we have used {\tt FeynRules 2.3} \cite{Alloul:2013bka}
to implement eq.~(\ref{form1}) together with the SM Lagrangian. 
Subsequently, {\tt Madgraph~v2.2.3} has
been utilised to compute the production cross-section $\sigma(pp\to \hx)$ through gluon
fusion and to calculate different partial decay widths of $\hx$.
In this study we have utilised $3.2~\fbi$ of ATLAS data at 13 TeV
to accommodate the observed resonance. 
In detail, we have used $\Delta N$, the discrepancy
between the observed and expected number of events $=13.6\pm3.69$. 
Further, for this purpose three $40$ GeV bins are chosen
for $690~{\rm GeV}\lsim \minv  \lsim 810$ GeV \cite{ATLAS-run-II-1}
with an efficiency of 0.4 \cite{CMS-run-II-2}.

In order to study the effect of BSM physics, we first show the  
variation of $\Gamma_X$ with changes in the new physics parameters,
$\kappa_g$ (left), $\kappa_A$ (right), in Fig.~\ref{fig:fit1}.
Here, we have varied $\kappa_g,\,\kappa_A$ in the span of $10^{-6}$ - $1$.
In these two plots the cyan coloured region represents the 
allowed $2\sigma$ range of $\Delta N$. 
The orange, golden and green coloured regions represent 
various zones in the $\Gamma_X$ - $\kappa_g~(\kappa_A)$ planes 
that are excluded from the 8 TeV LHC measurements of $\hx\to ZZ,\,Z\gamma$, $jj$
processes. The yellow coloured region
remains excluded from the measurement of $\hx\to \gamma\gamma$ \cite{ATLAS-diphoton}
process at the ATLAS with 8 TeV centre-of-mass energy.
Lack of precision measurements for the latter, assuming 
$\sigma(pp\to \hx)|_{13~\rm TeV}/\sigma(pp\to \hx)|_{8~\rm TeV}\approx 5$, predicts a $2\sigma$ upper 
bound \cite{Aad:2015mna} on $\sigma(pp\to\hx\to \gamma\gamma)|_{13~\rm TeV}$
inconsistent with the one observed with 13 TeV. We will discuss
this later in detail.
Finally, the gray coloured region remains excluded from the photon fusion process,
i.e., $\gamma\gamma\to\hx\to\gamma\gamma$, \cite{photon-photon} which predicts
a maximum for $Br(\hx\to\gamma\gamma)$, independent of 
$Br(\hx\to gg)$. The region excluded by the photon fusion
process is estimated by assuming that $\hx$ has 
only two decay modes $gg,\,\gamma\gamma$, i.e., 
$Br(\hx\to gg)$ + $Br(\hx\to \gamma\gamma) = 1$. The observed
limits on the $Brs$ are subsequently translated in terms
of $\kappa_g$ and $\kappa_A$.

It is evident from Fig.~\ref{fig:fit1}
that expecting $\Gamma_X$ as large as $45$ GeV or more is perfectly
consistent with the observed limits on $ZZ,\,\gamma\gamma,\,Z\gamma$
searches at the 8 TeV LHC. However, it is the di-jet searches
which rules out the region of parameter space with $\Gamma_X > 3$
GeV (right plot), corresponding to $\kappa_g \gsim \mathcal{O}(0.001)$ 
(left plot). The observed behaviour is well expected
as $\Gamma_{\hx\to gg}$ and thus, $\Gamma_X$ grows rapidly with $\kappa_g$ 
compared to that with $\kappa_A$, i.e., $\Gamma_{\hx\to \gamma\gamma}$
since the latter is suppressed by a factor of $\cos^4\theta_W/8$.
For $\kappa_A$ (estimated from $Br(\hx\to \gamma\gamma)$), the most stringent bound is 
coming from the photon fusion process which is represented
by the gray coloured region. 
For the photon fusion process, $Br(\hx\to\gamma\gamma)$ $\propto$ 
$1/\sqrt{\Gamma_X}$ \cite{photon-photon} and thus, {\it{smaller}} upper
bound on $Br(\hx\to\gamma\gamma)$ and hence, on $\kappa_A$
is expected for larger $\Gamma_X$.  This is evident from the right plot
of Fig.~\ref{fig:fit1}.
It is important to note that the 
photon fusion process can also provide an indirect bound
on $Br(\hx \to g g)$, i.e., on $\kappa_g$, assuming
$Br(\hx\to \gamma\gamma)+ Br(\hx \to g g)=1$. It is also apparent
that the photon fusion process discards $\Gamma_X \gtrsim$ 
$0.3$ GeV which is $10$ times smaller than the one predicted
from the di-jet search limit. Hence, given the observed large $\Gamma_X$
from the ATLAS, one needs almost the {\it equal} amount of $\Gamma_X$
from the hitherto unseen decay modes of this resonance, e.g., invisible decays.
Here, we use $\Gamma_X=\Gamma_{\hx\to \gamma\gamma}
+ \Gamma_{\hx\to ZZ}+ \Gamma_{\hx\to Z\gamma}+ \Gamma_{\hx\to jj}$,
as expected from eq.~(\ref{form1}), to estimate $Br(\hx\to \gamma\gamma)$ for 
the photon fusion process \cite{photon-photon}.
It is now clear that in the chosen setup, no realistic
values of $\kappa_A,\,\kappa_g$ parameters can account
for a total $\Gamma_X \gsim 0.3$ GeV. Thus, the presence of a {\it{huge}} additional
decay width is essential for the studied construction
which will be tightly constrained from the dark matter and monojet searches.

The discussion presented so far concerning the photon fusion process has
one caveat related to the estimation of $Br(\hx\to gg)$.
So far, we have used eq.~(\ref{form1}) to estimate
$\Gamma_X$, however, while evaluating the effect
of photon fusion process on $Br(\hx\to gg)$, i.e., on $\kappa_g$
(left plot of Fig.~\ref{fig:fit1}), we have used $Br(\hx\to \gamma\gamma)+Br(\hx\to gg)=1$
which is {\it apparently} contradicting. At this point one must note
that in the given construction the quantities $Brs(\hx\to Z\gamma,\,ZZ)$, as already explained,
are suppressed compared to $Br(\hx\to \gamma\gamma)$. Moreover, so far
we have no information available for processes like 
$\gamma Z,\,ZZ\to \hx$. Thus, the assumption
$Br(\hx\to gg)=1-Br(\hx\to \gamma\gamma)$ remains useful
for estimating the scale of $Br(\hx\to gg)$. Using all 
the available branching fractions instead would yield {\it weaker}
upper bounds on $Br(\hx \to gg)$, i.e., on $\kappa_g$.

It is evident from Fig.~\ref{fig:fit1} that $\Gamma_X \gsim 0.3$ GeV 
appears excluded from the relevant existing LHC limits and 
from the constraint of photon fusion process. This observation
demands the existence of {\it huge} additional decay width to 
reach the target of $45$ GeV. If we call this additional
width as $\Gamma^{add}_X$, without specifying the origin, then
one can write $\Gamma^{tot}_X \equiv \Gamma_X =$ $\Gamma_{\hx\to \gamma\gamma}
+ \Gamma_{\hx\to ZZ}$ $+ \Gamma_{\hx\to Z\gamma}+ \Gamma_{\hx\to jj}
+\Gamma^{add}_X$. This approach will modify all the associated
branching ratios as will be explored subsequently by {\it choosing}
three different values of the total decay widths: (1) $1$ GeV 
(small width), (2) $10$ GeV (moderate width) and (3) $45$ GeV (large width).

\begin{figure*}[t!]
\includegraphics[width=5.70cm,height=3.75cm,angle=0]{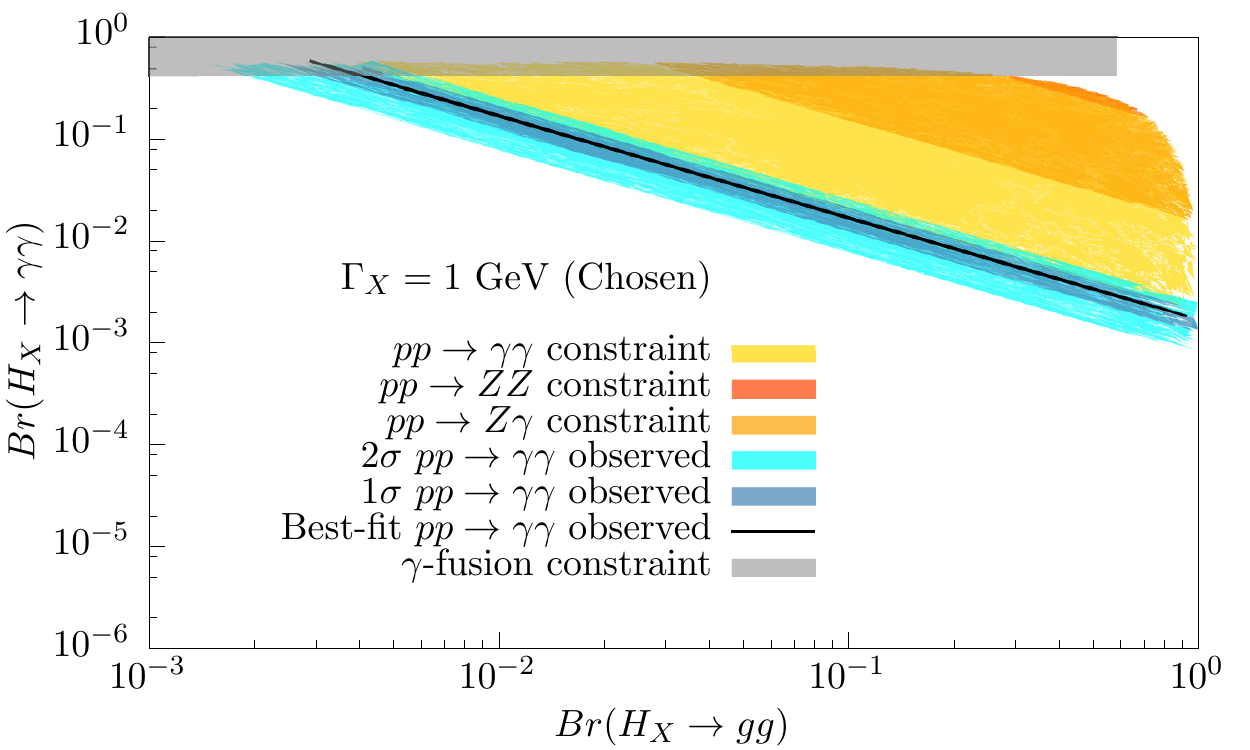}
\includegraphics[width=5.70cm,height=3.75cm,angle=0]{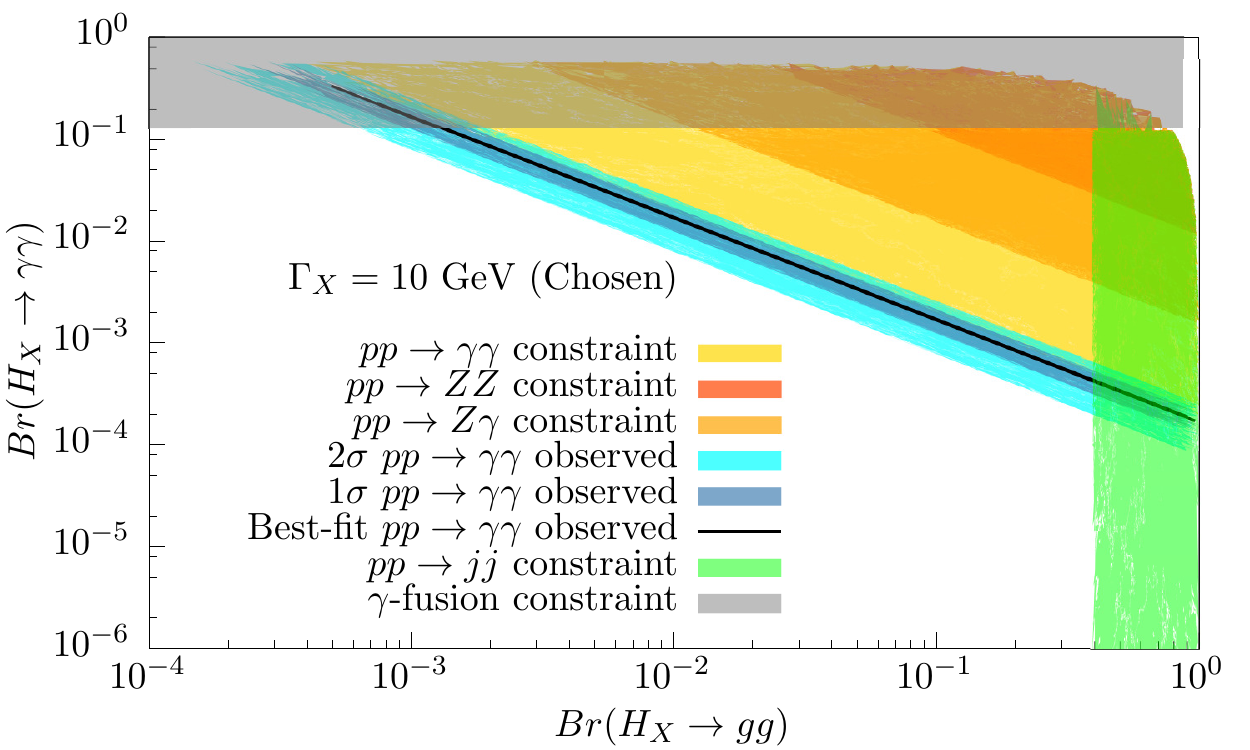}
\includegraphics[width=5.70cm,height=3.75cm,angle=0]{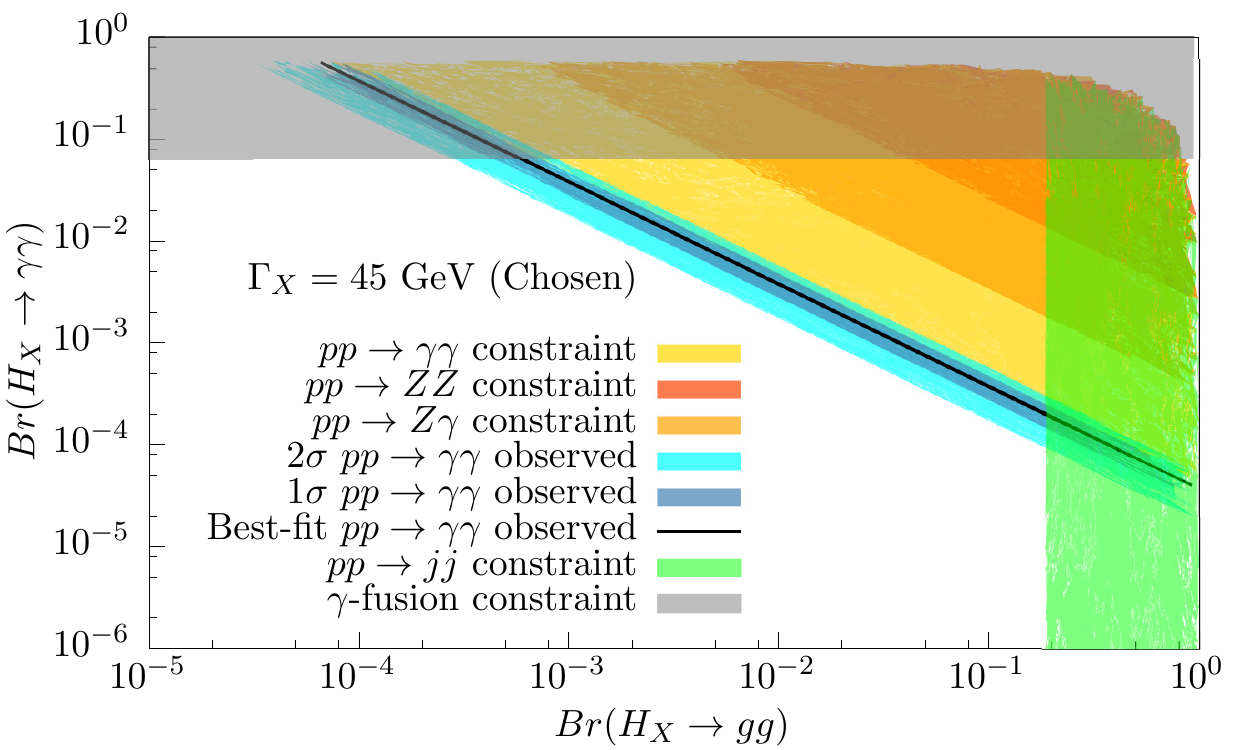}
\caption{Correlations in the $Br(\hx\rightarrow gg$) - $Br(\hx\rightarrow \gamma\gamma$) 
plane for the three different choices of $\Gamma_X$: 
1 GeV (left), 10 GeV (middle), and 45 GeV (right) compatible
with the observed $\sigma(pp\to\hx\to\gamma\gamma)$. 
The details are explained in the text.}
\label{fig:fit}
\end{figure*}

The subsequent effects of the aforesaid construction are 
explored in Fig.~\ref{fig:fit} where we have investigated
the impact of diverse LHC and photon fusion constraints 
in the $Br(\hx\to gg)$ - $Br(\hx\to \gamma\gamma)$ plane.
These two $Brs$ are expected to show {\it{some kind}} of correlation\footnote{This
correlation may disappear if this new resonance arises through the 
photon fusion process \cite{photon-photon}.}
between them since the observed excess appears through $gg\to\hx$ 
process followed by $\hx \to \gamma\gamma$ decay. It is
also possible to observe a similar correlation
in the $\kappa_g,\,\kappa_A$ plane since $Br(\hx\to gg),
\,Br(\hx\to \gamma\gamma)\propto \kappa^2_g,\,\kappa^2_A$, respectively.
In Fig.~\ref{fig:fit} the black coloured line represents the best-fit 
value corresponding to $\Delta N=13.6$ while the cyan and blue 
coloured bands represent the $1\sigma~(9.91 \lsim \Delta N \lsim 17.29)$ and 
$2\sigma~(6.22 \lsim \Delta N \lsim 20.98)$ allowed regions
in the concerned planes, respectively. 
The orange, golden, green and yellow coloured regions, similar
to Fig.~\ref{fig:fit1}, represent 
various zones in the concerned plane  
that are excluded from 8 TeV LHC limits on $\hx\to ZZ,\,Z\gamma$,
$jj$ and $\gamma\gamma$ processes.
In the case of $\hx \to \gamma\gamma$ process, assuming
$\sigma(pp\to \hx)|_{13~\rm TeV}/\sigma(pp\to \hx)|_{8~\rm TeV}\approx 5$,
one would expect a $2\sigma$ upper 
bound \cite{Aad:2015mna} on $\sigma(pp\to\hx\to \gamma\gamma)|_{13~\rm TeV}$
as $10$ fb using the ATLAS data. This is in tension with 
the 13 TeV ATLAS observation \cite{ATLAS-run-II-1} and rules out 
higher values of the observed $\sigma(pp\to\hx\to\gamma\gamma)$, 
starting from the central one.
A similar analysis using the CMS data \cite{CMS:2015cwa} excludes the higher 
values of the observed $\sigma(pp\to\hx\to \gamma\gamma)|_{13~\rm TeV}$ 
\cite{CMS-run-II-2} beyond $1\sigma$.

Lastly, the photon fusion process at the LHC, which predicts
a maximum for $Br(\hx\to\gamma\gamma)$ independent of 
$Br(\hx\to gg)$, rules out the gray coloured region in the $Br(\hx\to gg)$
 - $Br(\hx\to\gamma\gamma)$ plane. It is interesting
to note that the constraint for the photon fusion was derived
with the assumption of $Br(\hx\to gg)+ Br(\hx\to\gamma\gamma)=1$
which discards a region where $Br(\hx\to gg)$
+ $Br(\hx\to \gamma\gamma) > 1$.  For the three chosen
values of $\Gamma_X$, the maximum $Br(\hx\to \gamma\gamma)$
is estimated  \cite{photon-photon} as $\sim 0.42,\,0.13,\,0.06$, respectively and thus,
the regions with $Br(\hx\to gg)> 0.58$ (left plot of Fig.~\ref{fig:fit}),
$Br(\hx\to gg)> 0.87$ (middle plot of Fig.~\ref{fig:fit}),
$Br(\hx\to gg)> 0.94$ (right plot of Fig.~\ref{fig:fit}) remain
ruled out. The upper limits of $Br(\hx\to gg)$, as depicted
in Fig.~\ref{fig:fit} are purely illustrative.
This is because, following our earlier discussion, $Br(\hx\to gg)$
$=1-Br(\hx\to\gamma\gamma)$ estimated in a regime when 
$\Gamma^{add}_X\approx \Gamma^{tot}_X\equiv \Gamma_X$
appears simply illustrative.
For the rest of the processes
the primary productions are driven by the gluon fusion process. The latter
gives a high value for $Br(\hx\to gg)$ with increasing $\kappa_g$
and as a consequence remains excluded from the di-jet search limits,
especially for moderate to large $\Gamma_X$. For example,
for the choice of $\Gamma_X=10$ GeV one gets $Br(\hx\to gg)_{max}\sim 0.40$
(middle plot of Fig.~\ref{fig:fit}) while for 
the choice of $\Gamma_X=45$ GeV one ends up with $Br(\hx\to gg)_{max}\sim 0.20$
(right plot of Fig.~\ref{fig:fit}). In the case of small decay width (left plot of Fig.~\ref{fig:fit}) 
constraint from the di-jet searches remains ineffectual.

It is evident from eq.~(\ref{form1})
that $Br(\hx\to ZZ)$, $Br(\hx\to Z\gamma)$ are 
suppressed compared to $Br(\hx\to \gamma\gamma)$ by factors
of $\tan^4\theta_W$ and $\tan^2\theta_W$ (numerically $\sim 0.09$ and $0.3$),
respectively which is also apparent from Fig.~\ref{fig:fit}. 
Thus, unless one introduces interaction
like $\kappa_W W^a_{\mu\nu}W^{\mu\nu}_a$ ($W^a_{\mu\nu}$
as the $SU(2)$ field strengths) these modes remain sub-leading.
One can, nevertheless, compensate these deficits with a larger
$Br(\hx\to gg)$, assuming $gg\to \hx$ to be the leading production channel.
These behaviours are reflected in Fig.~\ref{fig:fit} where
the regions excluded from $ZZ$ and $Z\gamma$ searches
appear with lateral shifts towards larger $Br(\hx\to gg)$ values
compared to $Br(\hx\to \gamma\gamma)$ values, required
to reproduce the observed excess.
Larger $Br(\hx\to gg)$ and hence larger $\Gamma_X$
appear naturally for higher $\kappa_g$ values which are in tension
with the di-jet searches.  Increasing $\kappa_A$ receives 
constraint from the photon fusion process. 
The $ZZ$ and $Z\gamma$ constraints, as already mentioned, require large 
values for both of the $Br(\hx\to gg)$ and $Br(\hx\to \gamma\gamma)$.
The former faces tension from the di-jet search limits (moderate
and large $\Gamma_X$ scenarios)
while the latter, if not excluded by the photon fusion constraint,
might give larger $\Delta N$ than actually observed. 
Hence, the parameter space ruled out by these constraints 
do not affect the signal region compatible for explaining the observed excess.
The key feature of Fig.~\ref{fig:fit} is the prediction of
the value of product\footnote{The fact that the same product is $\sim \mathcal{O}
(10^{-11})$\cite{cernyellow} for a 750 GeV state with SM-like properties
justifies BSM nature of this excess.} $Br(\hx\to g g)\times Br(\hx \to \gamma\gamma)$ 
(henceforth written as $Br^2(\gamma\gamma\times gg)$).
From the best-fit line we observe that this value changes
from $\mathcal{O}(10^{-3})$ to $\mathcal{O}(10^{-5})$
as the chosen $\Gamma_X$ changes from $1$ GeV to $45$ GeV. 
Explicitly, $0.08(1.73)\times 10^{-3(5)}\lsim Br^2(\gamma\gamma\times gg)\lsim$
$1.93(4.28)\times 10^{-3(5)}$ for $\Gamma_X=1(45)$
GeV using $2\sigma$ limits on $\Delta N$. We will use 
these information subsequently.

Now we are ready to discuss the presence of other BSM
particles that are essential to explain this excess through
higher order processes. Information about these states
are encapsulated within $\kappa_g,\,\kappa_A$ (see eq.~(\ref{form1})).
These states must not be very heavy to avoid propagator suppression and at the same
time, must possess sizable couplings with $\hx$ to reproduce
the detected excess. Concerning the leading production, i.e., 
$gg \to \hx$, the possible candidate(s) is(are) either 
new coloured scalar(s) $\Phi$ or additional coloured
fermion(s) $F$, possibly vector-like.
These new particles must simultaneously couple to gluons as well as to $\hx$
and, are possibly embedded in a representation of
some larger symmetry group.
If these new scalars/fermions are also responsible for producing
an enhanced $Br(\hx\to\gamma\gamma)$, they must carry
electrical charges to get coupled to a photon.
However, the other non-minimal possibility is to consider 
another set of uncoloured but electrically charged 
fermion(s), scalar(s) or gauge boson(s) (appears
in theories with extended non-Abelian gauge sector). Note that 
contributions from new chiral fermions produce a 
destructive effect compared to the bosonic contributions
and thus, often are not compatible with the observed excess.
On the other hand, vector-like fermions remain
a viable alternative. The presence of an extended
scalar sector has additional phenomenological advantages,
e.g., stability of the SM-Higgs potential up to the Planck scale
\cite{Sher:1988mj,EliasMiro:2012ay,Alekhin:2012py,Buttazzo:2013uya}.
This argument also holds true for new gauge boson(s). 
We, however, do not consider
them in this article since they are hinted to be rather heavy 
$\gsim 2.5$ TeV \cite{ATLAS:2012ak, Khachatryan:2014dka}.
In a nutshell, we conclude that to accommodate the observed
di-photon excess one needs sizable couplings between $\hx$
and the new particles, for which coloured and/or electrically
charged scalars or fermions remain the realistic
options. Moreover, in the presence of the said new states,
an enhanced $Br(\hx\to\gamma\gamma)$ is more anticipated compared
to an enlarged $Br(\hx\to gg)$ as for the latter experimental
evidences are still missing.

In the presence of a new BSM scalar $\Phi$, with mass $M_\Phi$, electric charge $Q_\Phi$ 
(in the units of $|e|$) and number of colour $N^c_\Phi$, the 
$Br(\hx\to \gamma\gamma)$ can be written as \cite{Carena:2012xa,Jaeckel:2012yz}:
{\small
\beq
\label{form2}
Br({\rm H}_{\rm X} \to \gamma \gamma)=\frac{\alpha^2 M^3_X}{1024\pi^3 \Gamma_X}
\Big |\frac{g_{\Phi\Phi {\rm H}_{\rm X}}}{M_{\Phi}^2} N_\Phi^c Q_{\Phi}^2 A_{\Phi}(x_{\Phi}) \Big |^2.
\eeq}
Here, $\alpha_{em}$ is the electromagnetic coupling constant, 
$g_{\Phi\Phi\hx}$ represents the coupling between $\Phi$ and $\hx$
and the detail of $A_{\Phi}(x_{\Phi})$ function, where $x_\Phi=4 M^2_\Phi/M^2_X$, 
is given in Ref.~\cite{Carena:2012xa}. Keeping in mind the issue
of perturbativity we choose $-\sqrt{4\pi}\lsim g_{\Phi\Phi\hx} \lsim \sqrt{4\pi}$,
in our numerical analyses. The quantity $g_{\Phi\Phi\hx}$
parametrises the information about the vacuum expectation value (VEV) of $\hx$
and the amount of {\it possible} mixing between $\hx$ and the SM-Higgs.
From eq.~(\ref{form2}) it appears that a larger $g_{\Phi\Phi{\rm H}_{\rm X}}$ 
is useful to produce a bigger $Br(\hx\to \gamma\gamma)$.
In reality, however, such scenarios are unrealistic
as they correspond to either experimentally challenging large mixing within
$\hx$ and the SM-Higgs or a large VEV for $\hx$ inconsistent
with the electroweak precision tests \cite{Agashe:2014kda}.

\begin{figure}[t!]
\includegraphics[width=4.45cm,height=3.7cm]{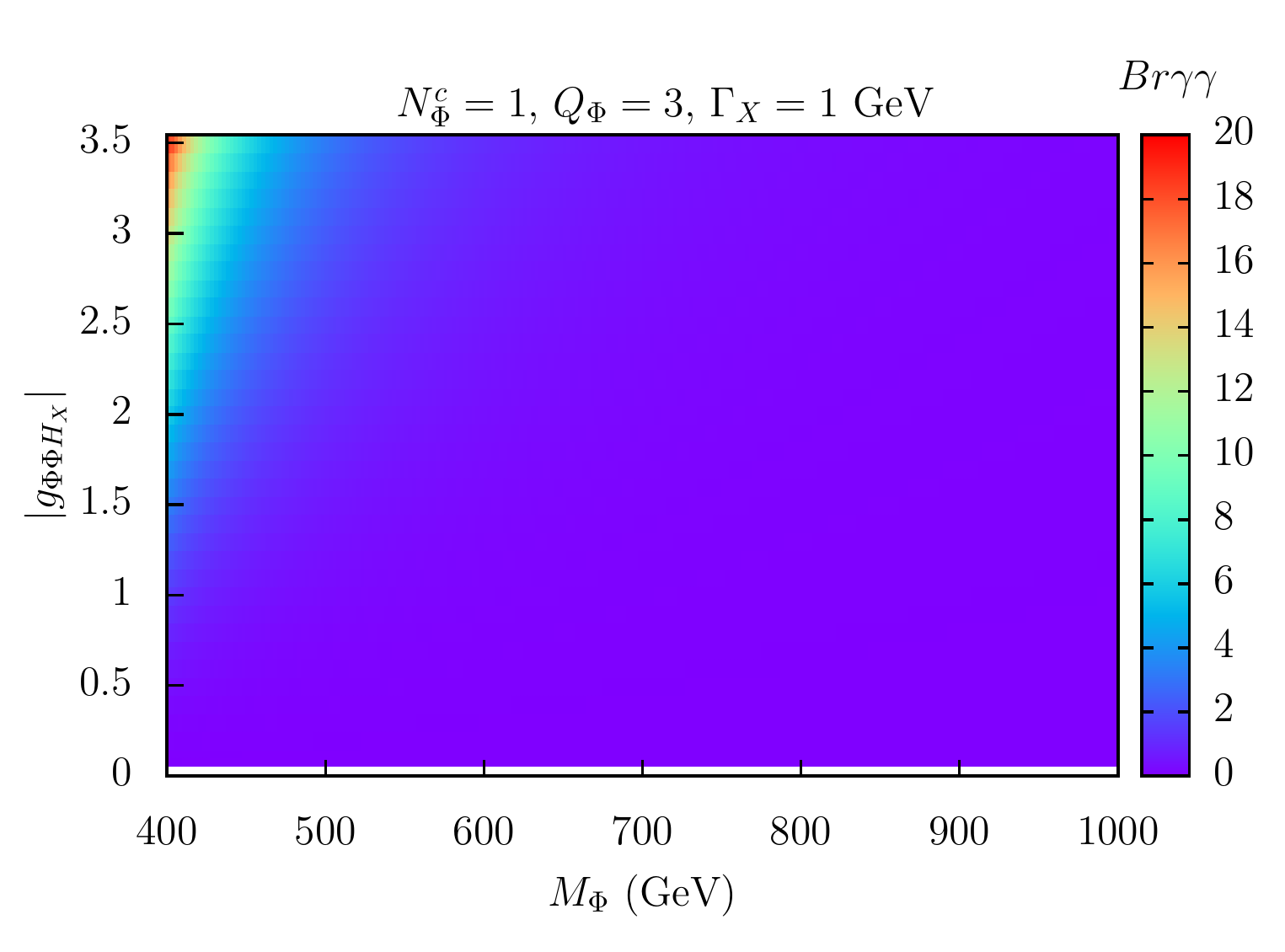}
\hspace*{-0.2cm}
\includegraphics[width=4.45cm,height=3.7cm]{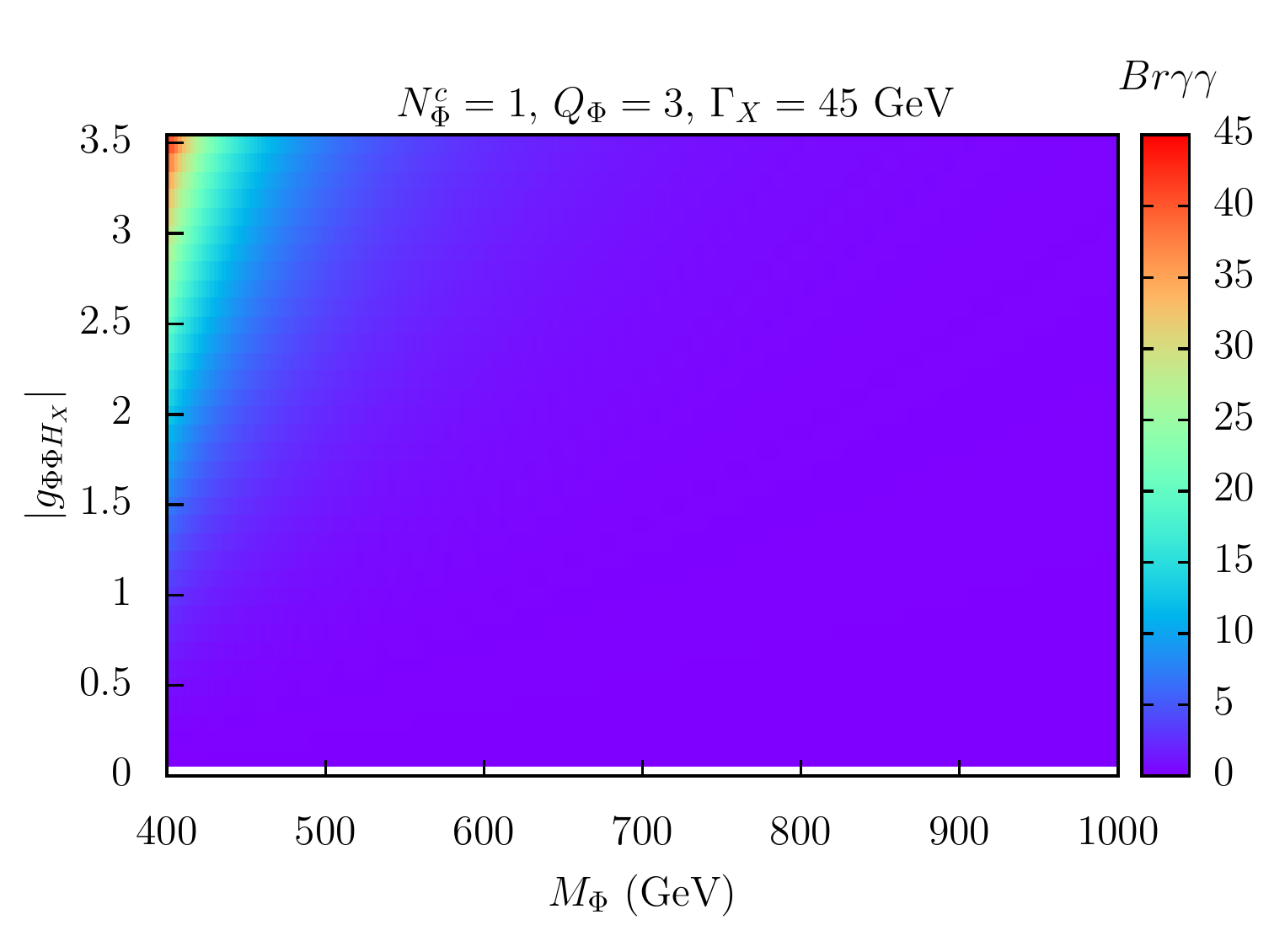}
\caption{Plots showing the variation of $Br\gamma\gamma
= Br(\hx\rightarrow\gamma\gamma)\times 10^n $ in the ${M_{\Phi}}$ - 
$|g_{\Phi\Phi {\rm H}_{\rm X}}|$ plane for $N_{\Phi}^c,\,Q_{\Phi}=1,\,3$ with 
$\Gamma_X=$ $1$ GeV (left) and 45 GeV (right). The chosen ranges
for $M_\Phi$ and $g_{\Phi\Phi {\rm H}_{\rm X}}$ are explained in the text.
The multiplicative factor $n=9(11)$ for $\Gamma_X=1(45)$ GeV.}
\label{fig:scalar_contrb}
\end{figure}

It is apparent from eq.~(\ref{form2}) that depending
on the values of $M_\Phi,\,Q_\Phi$ and $N^c_\Phi$, the quantity $Br(\hx\to\gamma\gamma)$
can receive sizable enhancement. An enlargement is also possible
if the future LHC observation confirms a smaller $\Gamma_X$.
In our numerical analyses we choose $400~{\rm GeV}\lsim M_{\Phi} \lsim 1000~{\rm GeV}$,
consistent with the existing collider bounds on such 
exotic particles \cite{Chatrchyan:2012ya,ATLAS:2012hi,ATLAS:2014kca}.
A sample variation of $Br(\hx\to\gamma\gamma)$ in the $M_\Phi-g_{\Phi\Phi{\rm H}_{\rm X}}$
plane for a colour singlet ($N^c_\Phi=1$), triply charged $(Q_\Phi=3)$,
scalar with different $\Gamma_X$, 1 GeV (left) and 45 GeV (right)
is shown in Fig.~\ref{fig:scalar_contrb}. 
%
It is evident from Fig.~\ref{fig:scalar_contrb} that an experimentally
viable light, i.e., $M_\Phi=400$ GeV, colour singlet
$\Phi$ with $Q_\Phi=3$ can produce at most a $Br(\hx\to \gamma\gamma)$
$\sim \mathcal{O}(10^{-8})$ (left plot) when $\Gamma_X$ is small, i.e., 1 GeV. Choosing
$\Gamma_X=45$ GeV instead one faces a reduction by a factor of $45$ (right plot).
From eq.~(\ref{form2}) we see that $Br(\hx\to \gamma\gamma)\propto Q^4_\Phi$.
Thus, even for an exotic colour singlet $\Phi$ with $Q_\Phi=10$, one would
expect a maximum $Br(\hx\to\gamma\gamma)\sim \mathcal{O}(10^{-6})$
keeping $\Gamma_X,\,M_\Phi=1~{\rm GeV},\,400$ GeV. Now, from our 
previous discussion in the context of Fig.~\ref{fig:fit}, we 
have estimated $Br^2(\gamma\gamma\times gg)$ as $\sim \mathcal{O}(10^{-3})$
and $\sim \mathcal{O}(10^{-5})$ for the choice of $\Gamma_X=1$ and 45 GeV, respectively
from the best-fit value of $\Delta N$. Hence, the maximum
$Br(\hx\to\gamma\gamma)$, extracted from Fig.~\ref{fig:scalar_contrb}
using eq.~(\ref{form2}) for a 400 GeV $\Phi$
with $N^c_\Phi,\,Q_\Phi=1,\,10$, would give an {\it unrealistic} $Br(\hx\to gg)\sim $
$400(180)$ for $\Gamma_X=1(45)$ GeV scenarios.
One may try to consider a
{\it similar} but {\it coloured} (say $N^c_\Phi =3$) $\Phi$
which predicts a maximum $Br(\hx\to\gamma\gamma)\sim \mathcal{O}$
$(10^{-5})$ for $\Gamma_X=1$ GeV. However, one still needs 
an {\it unrealistic} $Br(\hx\to gg)\sim 50$ in this scenario.
Moreover, for a $\Phi$ with non-zero colours one must carefully investigate the 
$\hx\to jj$ constraint, even for a realistic $Br(\hx\to gg)$, 
especially for moderate to large $\Gamma_X$.

\begin{figure}[t!]
\includegraphics[width=4.45cm,height=3.7cm]{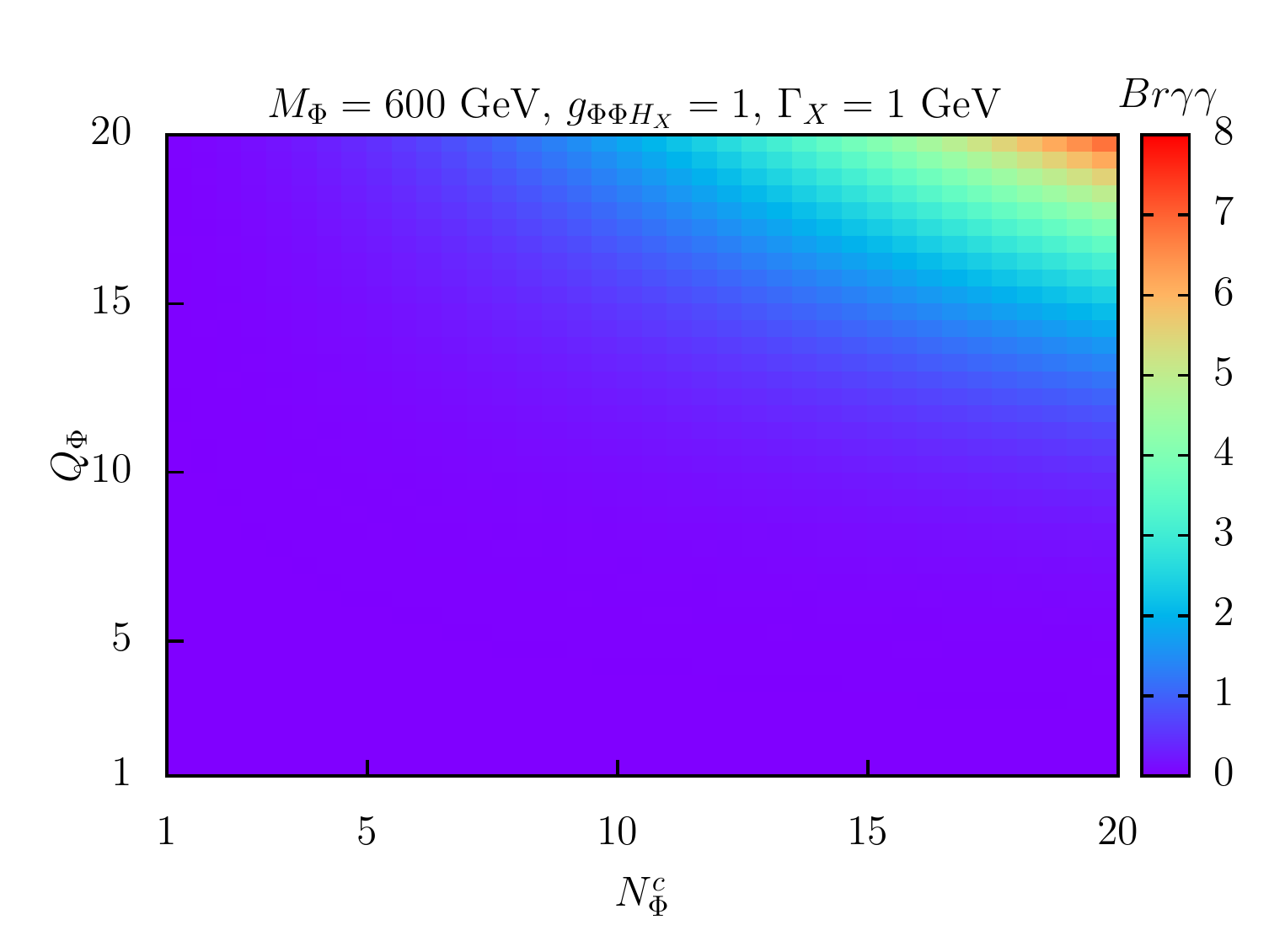}
\hspace*{-0.2cm}
\includegraphics[width=4.45cm,height=3.7cm]{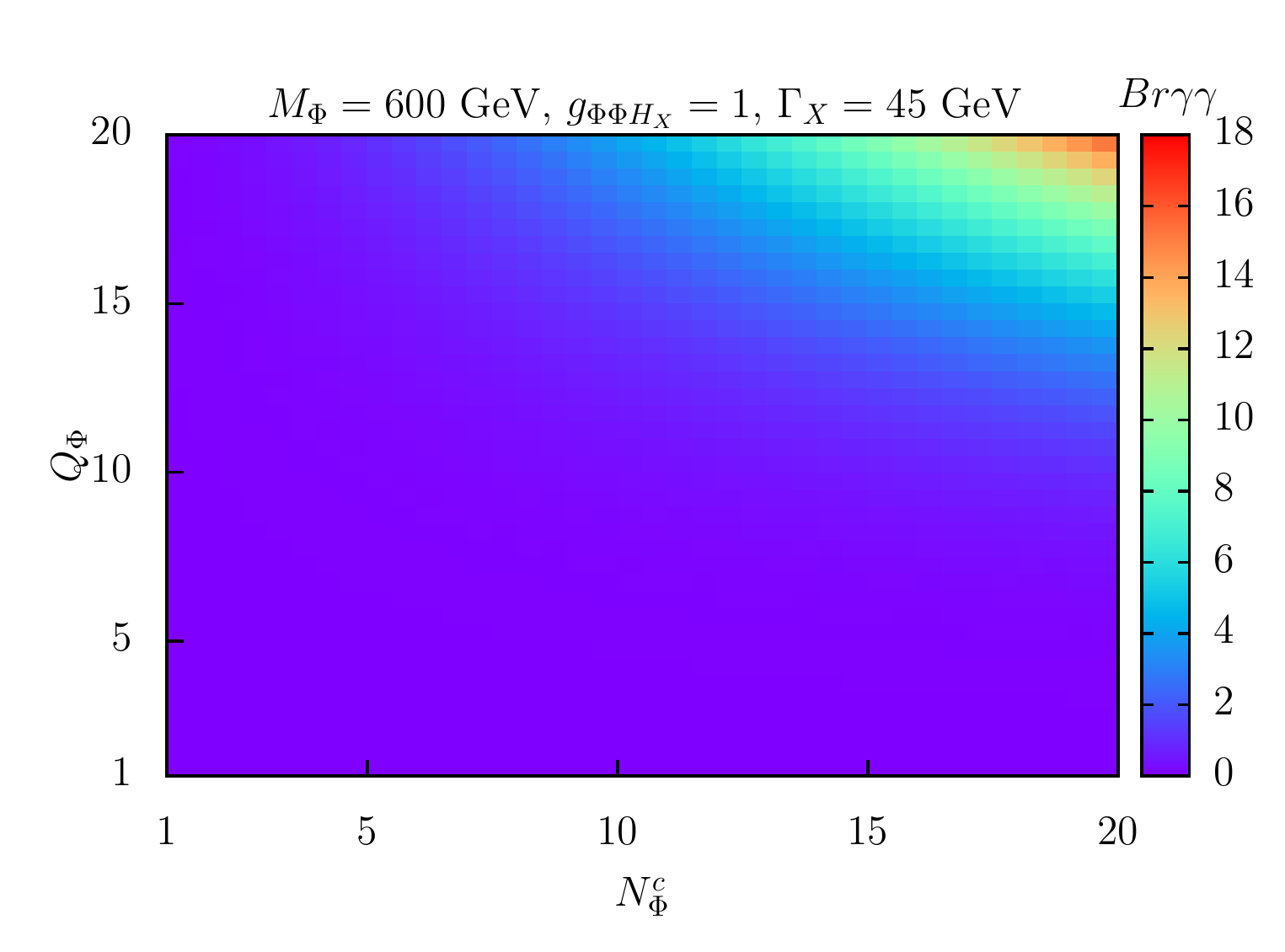}
%
\caption{Variation of $Br\gamma\gamma=Br(\hx\rightarrow\gamma\gamma)\times 10^n$ in the 
$N_{\Phi}^c$ - $Q_{\Phi}$  plane for $M_{\Phi}=$ 600 GeV, $g_{\Phi\Phi {\rm H}_{\rm X}}=1$, 
$\Gamma_X=$ 1 GeV(left) and 45 GeV(right). Here, $n=5(7)$ for $\Gamma_X=$ 1(45) GeV.}
\label{fig:scalar_contrb2}
\end{figure}
From the last discussion it appears that the use of new 
BSM scalar is not adequate to explain the observed
excess. In order to explore this further we have plotted
the change of $Br(\hx\to\gamma\gamma)$ in the $N^c_\Phi$-$Q_\Phi$
plane in Fig.~\ref{fig:scalar_contrb2} with $M_\Phi,\,g_{\Phi\Phi {\rm H}_{\rm X}}=600$ GeV, 1
for the choice of $\Gamma_X=1$ GeV (left) and 45 GeV (right).
Here, we vary both $N^c_\Phi,\,Q_\Phi$ in the range of $1:20$
and the chosen values of $M_\Phi,\,g_{\Phi\Phi {\rm H}_{\rm X}}$
are purely illustrative. It is apparent from both of these plots
that to satisfy $Br^2(\gamma\gamma\times gg)$, consistent
with the observation of Fig.~\ref{fig:fit}, one should have
an {\it unrealistic} $Br(\hx\to gg)\sim 10(5)$ for  
$\Gamma_X=1(45)$ GeV. Adopting smaller $M_\Phi$ (say 400 GeV)
simultaneously with a larger $g_{\Phi\Phi {\rm H}_{\rm X}}$
(say $\pm 3$) one can reach a maximum $Br(\hx\to \gamma\gamma)$
$\sim 0.012$ and $\sim 0.00025$
for $\Gamma_X=1$ and $45$ GeV, respectively considering
$N^c_\Phi \gsim14,\,Q_\Phi\gsim 16$. Here, we have used eq.~(\ref{form2})
and information from Fig.~\ref{fig:scalar_contrb2}. So, apparently these 
exotic scenarios can give a {\it realistic} $Br(\hx\to gg)\lsim \mathcal{O}(0.1)$,
consistent with the di-jet searches (see Fig.~\ref{fig:fit}).
However, this moderate $Br(\hx\to gg)$ value
may get excluded from the future LHC searches with expected higher 
sensitivity. Moreover, one must carefully re-evaluate
the maximum value for $g_{\Phi\Phi {\rm H}_{\rm X}}$
in a consistent theory framework. It is now evident
from the last discussion that the presence of BSM $\Phi$s, 
instrumental to reproduce the observed excess,
requires really high electric and colour charges.
Particles with such high colour charges
are expected to be produced amply at the LHC, unless
very massive and hence, rather stringent constraints
are expected on their existence. We thus, leave our 
discussion about the BSM scalars without further detail. We note in passing
that $Q_\Phi$ value as high as 20 can be 
interpreted as an effective electric charge,
keeping $N^c_\Phi$ fixed. For example data-set 
with $Q_\Phi=20$ for a fixed $N^c_\Phi$, using eq.~(\ref{form2}), can be 
thought of as a coloured/uncoloured multiplet with members of
{\it almost the same} masses and having electric charges
from $\pm 1$ to $\pm10$.

Let us now investigate a similar scenario in the presence
of new BSM vector-like fermion, $F$. For a fermion with mass $M_F$, electric charge $Q_F$
(in the units of $|e|$), number of colours $N^c_F$, the 
quantity $Br(\hx\to \gamma\gamma)$ is expressed as \cite{Carena:2012xa,Jaeckel:2012yz}:
{\small
\beq
\label{form3}
Br({\rm H}_{\rm X} \to \gamma \gamma)=\frac{\alpha^2 M^3_X}{1024\pi^3\Gamma_X}
\Big |\frac{2g_{FF{\rm H}_{\rm X}}}{M_{F}}N_{F}^c Q_{F}^2 A_{F}(x_{F}) \Big |^2.
\eeq}
Here, $g_{FF{\rm H}_{\rm X}}$ represents the generic coupling
between $F$ and $\hx$. The function $A_{F}(x_{F})$,
with $x_F=4M^2_F/M^2_X$, is given in Ref.~\cite{Carena:2012xa}.
We consider $500~{\rm GeV}\lsim M_F\lsim 1$ TeV
(see Ref.~\cite{BSM-fermion}
and references therein) while $g_{FF{\rm H}_{\rm X}}$ is varied in a range similar
to $g_{\Phi\Phi {\rm H}_{\rm X}}$, based on the same argument.

\begin{figure}[t!]
\includegraphics[width=4.45cm,height=3.7cm]{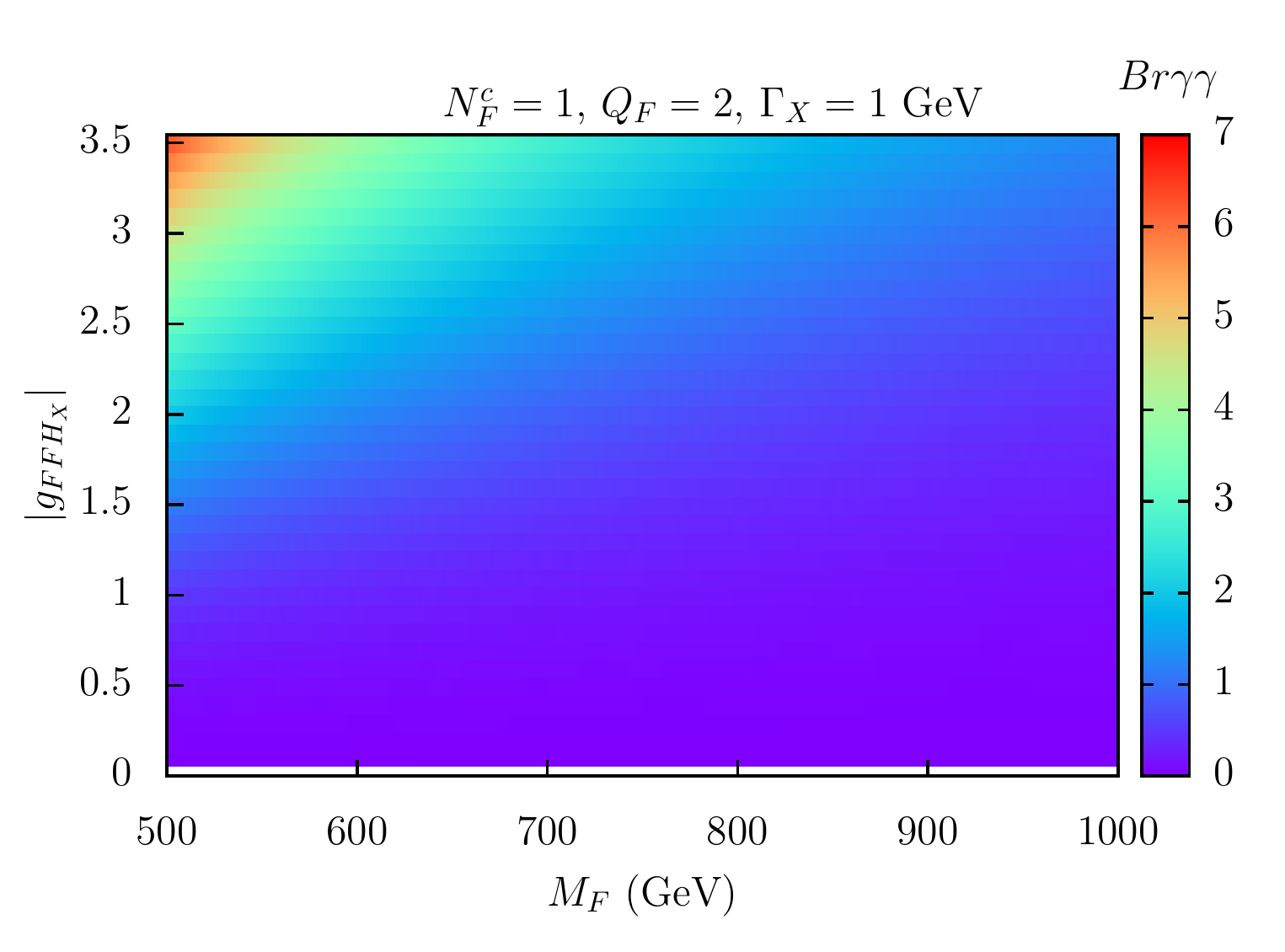}
\hspace*{-0.2cm}
\includegraphics[width=4.45cm,height=3.7cm]{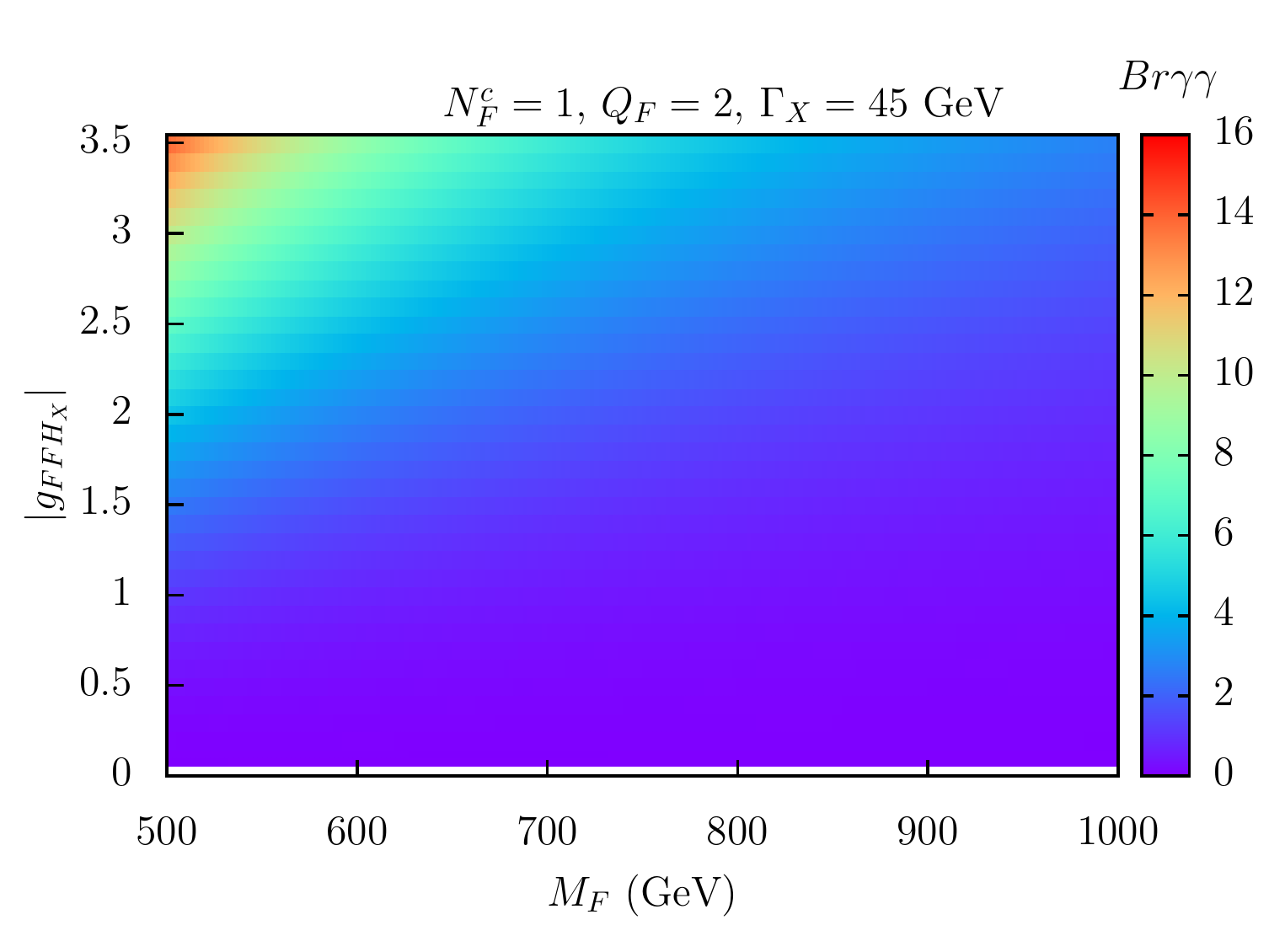}
%
\caption{Plots showing the variation of $Br\gamma\gamma=
Br(\hx\rightarrow\gamma\gamma)\times10^n$ in the ${M_{F}}$ - 
$|g_{FF {\rm H}_{\rm X}}|$ plane for $N_{F}^c,\,Q_{F}=1,\,2$ with 
$\Gamma_X=$ $1$ GeV (left) and 45 GeV (right). The chosen ranges
for $M_F$ and $g_{FF {\rm H}_{\rm X}}$ are explained in the text.
Here, $n=3(5)$ for the left(right) plot.} 
\label{fig:fermion_contrb}
\end{figure}
The sample variation of $Br(\hx\to \gamma\gamma)$ in the $M_F$ - $g_{FF{\rm H}_{\rm X}}$
plane for a colour singlet doubly charged ($Q_F=2$) fermion
is shown in Fig.~\ref{fig:fermion_contrb} for $\Gamma_X=1$
(left) and 45 GeV (right). It is easy to see from 
these plots that the presence of BSM fermions is more 
efficient to raise $Br(\hx\to \gamma\gamma)$ compared to the BSM scalars.
For example a colour singlet doubly charged fermion
can produce $Br(\hx\to\gamma\gamma)$ as high as $0.007$
and $\sim 10^{-4}$ for $\Gamma_X=1$ and 45 GeV, respectively.
These numbers are orders of magnitude larger compared
to the same from Fig.~\ref{fig:scalar_contrb} and, as stated before, can only
be achieved for a $\Phi$ with very high $Q_\Phi$ and $N^c_\Phi$.
These enhanced $Br(\hx\to\gamma\gamma)$ values are also
useful to estimate {\it realistic} values of $Br(\hx\to gg)$,
using the information from Fig.~\ref{fig:fit}. As an example,
from Fig.~\ref{fig:fermion_contrb}, with the maximum of $Br(\hx\to\gamma\gamma)$, 
one can estimate $Br(\hx\to gg)\sim 0.14(0.063)$ for $\Gamma_X=1(45)$ GeV 
using the derived bound on $Br^2(\gamma\gamma\times gg)$. 
Clearly, one can easily reproduce the observed excess, especially for 
smaller $\Gamma_X$, without any difficulty. However, for larger
$\Gamma_X$, depending on its value, some of
the $Br(\hx\to gg)$ values remain excluded from the di-jet searches
as already depicted in Fig.~\ref{fig:fit}.

\begin{figure}[t!]
\includegraphics[width=4.45cm,height=3.7cm]{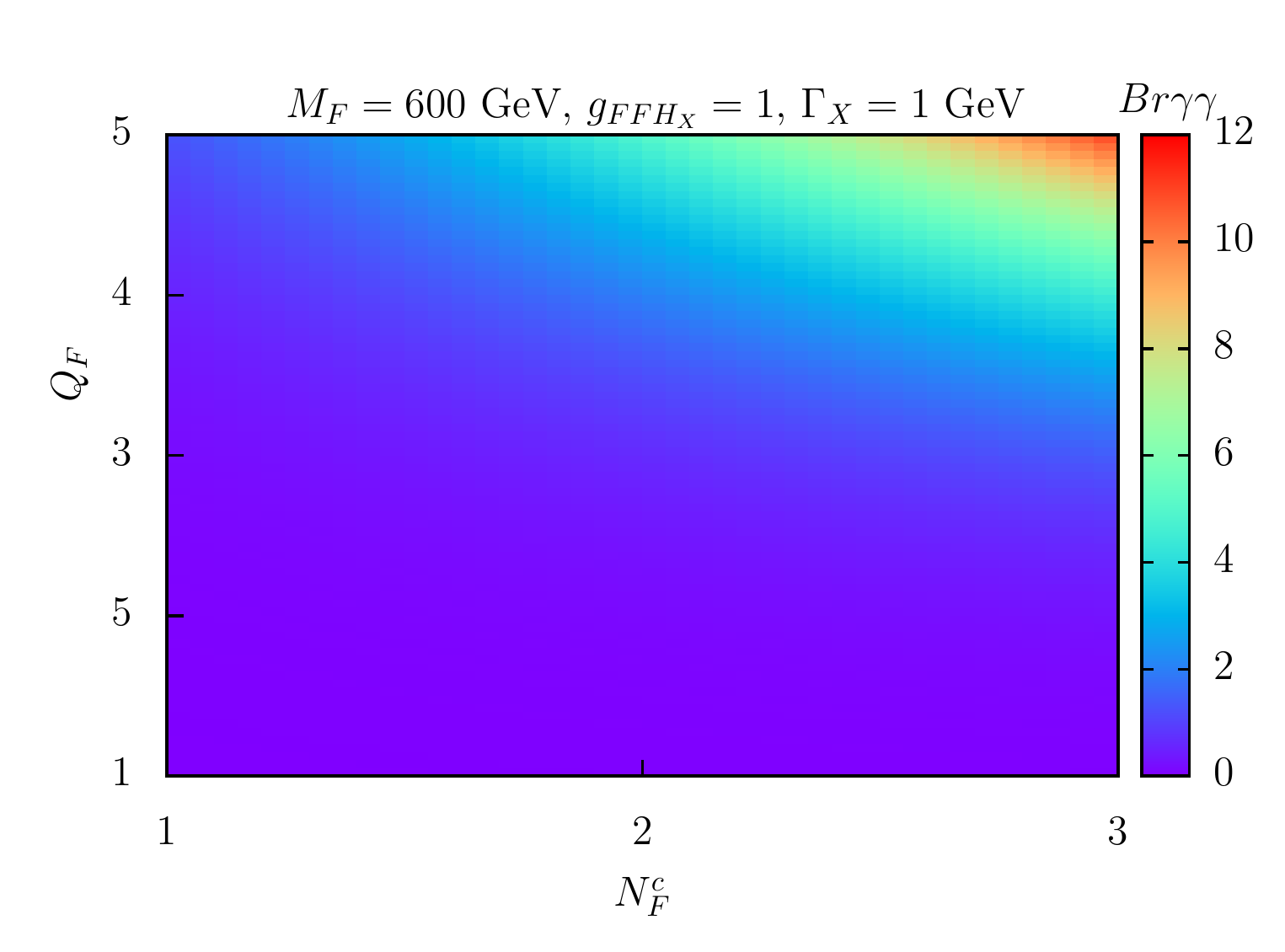}
\hspace*{-0.2cm}
\includegraphics[width=4.45cm,height=3.7cm]{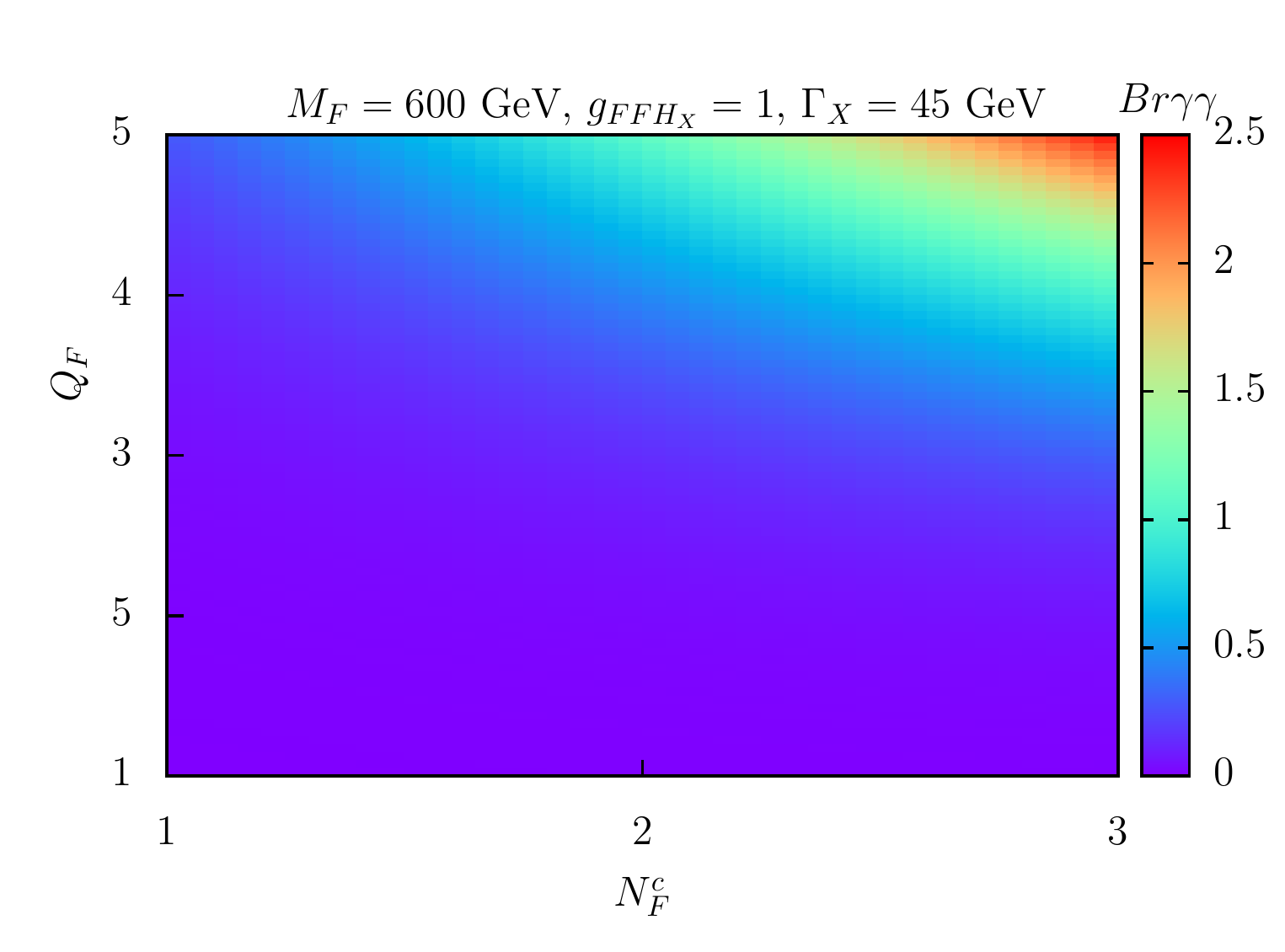}
%
\caption{Variation of $Br\gamma\gamma=Br(
\hx\rightarrow\gamma\gamma)\times 10^n$ in the 
$N_{F}^c$ - $Q_{F}$  plane for $M_{F}=$ 600 GeV, $g_{FF{\rm H}_{\rm X}}=1$, 
$\Gamma_X=$ 1 GeV(left) and 45 GeV(right). For the 
left(right) plot $n=2(3)$.}
\label{fig:fermion_contrb2} 
\end{figure}

In order to study the behaviour of $Br(\hx\to\gamma\gamma)$ 
with changes in the $N^c_F$-$Q_F$ values we have plotted the same 
in Fig.~\ref{fig:fermion_contrb2} for the choice of $\Gamma_X=1$ GeV (left) 
and 45 GeV (right). The chosen $M_F$, $g_{FF {\rm H}_{\rm X}}$ values
are purely illustrative. We vary
$N^c_F(Q_F)$ in the range of $1:3(5)$.
It is visible from these plots that an electrically
charged coloured fermion can generate $Br(\hx\to\gamma\gamma)$
as high as $0.12(0.0025)$ for $\Gamma_X=1(45)$ GeV.
Using the information from Fig.~\ref{fig:fermion_contrb2} and 
our knowledge of Fig.~\ref{fig:fit}, these numbers
predict $0.01(0.004)\lsim Br(\hx\to gg)\lsim 0.1(0.05)$ for $\Gamma_X=1(45)$ GeV.
These numbers are absolutely consistent with the di-jet searches
as shown in Fig.~\ref{fig:fit}. For exotic fermions
a lower value of $M_F$ (say 400 GeV) and a higher $g_{FF{\rm H}_{\rm X}}$ value
(say $\pm \sqrt{4\pi}$), especially for small $\Gamma_X$,
will produce $Br(\hx\to \gamma\gamma)$ either $>1$ or
inconsistent with the constraint of photon fusion process for certain region 
of the $N^c_F-Q_F$ plane, e.g., $1\lsim N^c_F\lsim 3,\,Q_F> 3$.
Further, compared to the BSM scalars, the BSM fermions
can appear instrumental to reproduce the observed
excess with realistic values of $Br(\hx\to gg)$
and $Br(\hx\to \gamma\gamma)$ even with $M_F=1$ TeV. As
for the latter with $g_{FF{\rm H}_{\rm X}}=1,\,\Gamma_X=$
$45$ GeV and with $1\lsim N^c_F(Q_F)\lsim 3(5)$
one gets $0.0001\lsim Br(\hx\to \gamma\gamma)\lsim 0.0008$
and thus, $0.0125\lsim Br(\hx\to g g)\lsim 0.1$,
consistent with Fig.~\ref{fig:fit}.
We note in passing that, similar to the scalars, one 
can also explain Fig.~\ref{fig:fermion_contrb2}, say $N^c_F,Q_F=1,\,4$, with 
a single uncoloured multiplet with quasi-degenerate masses for the members
and $Q_F$ ranging from $\pm 1$ to $\pm 3$.

The proficiency of the BSM fermions over the scalars
are now established. Although one can reproduce the 
excess with a colour singlet fermion with high $Q_F$ (see Fig.~\ref{fig:fermion_contrb2}),
nevertheless, it is an absolute necessity to explore
the scenario with $N^c_F >1$ as otherwise the 
expected BSM origin for $gg\to \hx$ process remains unexplained.
This scenario may receive constraint from
di-jet searches provided the enhanced efficiency expected from the future LHC operation.

\begin{figure}[t!]
\includegraphics[width=4.45cm,height=3.75cm]{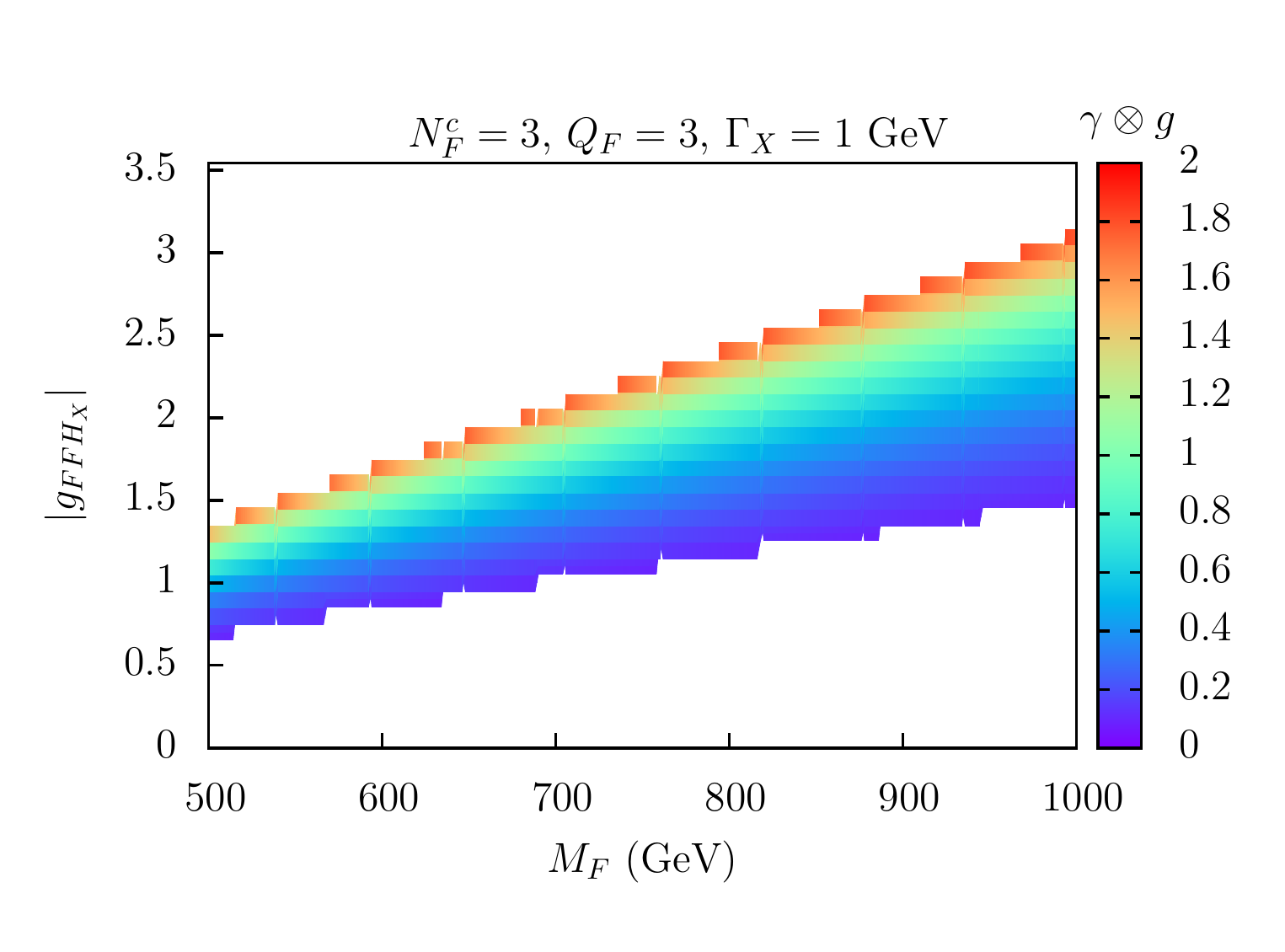}
\hspace*{-0.2cm}
\includegraphics[width=4.45cm,height=3.75cm]{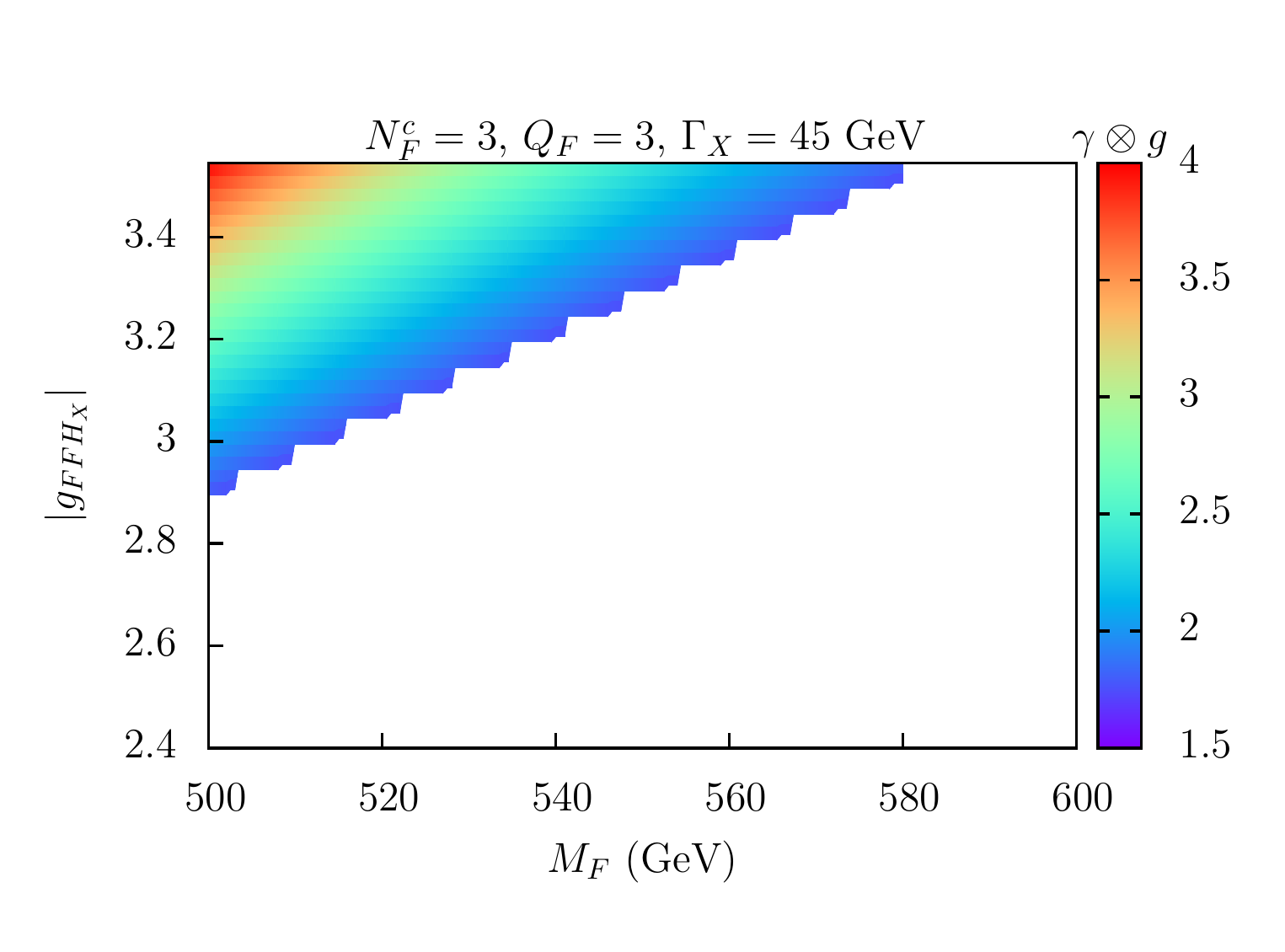}
%
\caption{Plots showing the variation of $\gamma\otimes g \equiv Br(\hx\to\gamma\gamma)\times$
$Br(\hx\to gg)\times 10^n$ in the ${M_F}-|g_{FF {\rm H}_{\rm X}}|$ plane
for $N^c_F=3,\,Q_F=3$ with $\Gamma_X=1$ GeV (left)
and 45 GeV (right). Here, $n=3(5)$ 
for $\Gamma_X=1(45)$ GeV.} 
\label{fig:fermion_contrb3}
\end{figure}

The exotic fermions, similar to $Br(\hx\to \gamma\gamma)$ 
(see eq.~(\ref{form3})), can also 
contribute to $Br(\hx\to gg)$. At the leading order this branching ratio  
\cite{Spira:1995rr} is given as:
{\small
\beq
\label{h2gg_fermion}
Br({\rm H}_{\rm X} \to gg)=\frac{\alpha_s^2 M^3_X}{512\pi^3\Gamma_X}
\Big |\frac{2g_{FF{\rm H}_{\rm X}}}{M_{F}} A_{F}(x_{F}) \Big |^2.
\eeq}
Here, $\alpha_s$ is the strong coupling constant. In our numerical
analysis we have multiplied $Br(\hx\to gg)$, as shown in eq.~(\ref{h2gg_fermion}),
by a factor of $1.5$, relevant for the higher order effects of strong interactions.
Using eqs.~(\ref{form3}) and (\ref{h2gg_fermion}) simultaneously
we have studied a sample variation of $Br^2(\gamma\gamma\times gg)$
in the $M_F$-$g_{FF{\rm H}_{\rm X}}$ plane as shown by 
Fig.~\ref{fig:fermion_contrb3} with $\Gamma_X=1$ GeV (left)
and 45 GeV (right). For this figure $M_F$, $g_{FF{\rm H}_{\rm X}}$
are varied as of Fig.~\ref{fig:fermion_contrb} and we work with
$N^c_F=3$ and $Q_F=3$.
The observed behaviours of $Br^2(\gamma\gamma\times gg)$ with 
different parameters, i.e., $M_F,\,g_{FF {\rm H}_{\rm X}}$
and $\Gamma_X$ are expected from eqs.~(\ref{form3}) and (\ref{h2gg_fermion}).
For example, both $Br(\hx\to \gamma\gamma)$ and $Br(\hx\to gg)$
are $\propto \Gamma^{-1}_X$ and thus, shrinking of the allowed parameter space, 
compatible with the observed excess, for larger $\Gamma_X,\,45$ GeV,
(right plot of Fig.~\ref{fig:fermion_contrb3}) is
anticipated. At the same time, these two branching ratios
are $\propto g^2_{FF {\rm H}_{\rm X}}/M^2_F$ (see eqs.~(\ref{form3}),\,(\ref{h2gg_fermion})).
Hence, apparent lowering of $Br^2(\gamma\gamma\times gg)$
for larger $M_F$ values must be compensated with 
larger $g_{FF {\rm H}_{\rm X}}$ values in order to remain compatible with the excess.
This feature is depicted in Fig.~\ref{fig:fermion_contrb3},
notably for the left one. The most useful  
aspect of Fig.~\ref{fig:fermion_contrb3} is connected
with the estimation of future detection possibility
for the process $gg \to H^*_X \to F \bar{F}$. Assuming that 
the future measurements indicate a narrow width
for this excess, say 1 GeV, then the room for 
measuring $\sigma(gg \to H^*_X \to F \bar{F})$ is less promising for 
two reasons: (1). The expected enhancement in the production 
for low $M_F$ region is ameliorated with a 
relatively small $g_{FF {\rm H}_{\rm X}}$ and (2). 
In the high $g_{FF {\rm H}_{\rm X}}$ regime, 
the same logic remains applicable through heavier $M_F$. These two features are visible
from the left plot of Fig.~\ref{fig:fermion_contrb3}.
On the contrary, a more stringent limit, i.e., $Br^2(\gamma\gamma\times gg)
\sim \mathcal{O}(10^{-5})$, for larger $\Gamma_X=45$ GeV 
prefers smaller $M_F$ and larger $g_{FF {\rm H}_{\rm X}}$
(see right plot of Fig.~\ref{fig:fermion_contrb3}). Both 
of these would appear useful to enhance $\sigma(gg \to H^*_{X} \to F \bar{F})$.

{\it Conclusions:} To summarise, the LHC run-II has already observed
an excess in the di-photon invariant mass distribution near 750 GeV. This
excess, as argued in this article, definitely requires
BSM physics. In this article we tried to explore this excess, assuming a spin-0 nature, 
using an simplified effective Lagrangian, sensitive to {\it new} physics effects. 
The chosen framework helped us to estimate a lower bound of $\Gamma_X$,
consistent with the different LHC constraints and photon fusion process,
for changes in the new physics parameters, $\kappa_g,\,\kappa_A$.
We have also explored the possible
correlation between $Br(\hx\to \gamma\gamma)$ and 
$Br(\hx\to gg)$ in the light of the observed excess
and diverse possible constraints. This correlation provides a {\it model-independent 
but $\Gamma_X$-dependent} bound on $Br(\hx\to \gamma\gamma)\times$ 
$Br(\hx\to gg)$.
Subsequently, we have utilised this correlation to scrutinise 
the effect of other BSM scalars, fermions with
various electric charge, number of colour which simultaneously
couple to $\hx$ and $gg,\,\gamma\gamma$ and might 
appear instrumental to reproduce this excess through higher
order processes. Our analyses show that to accommodate
the observed excess, the presence of additional BSM fermions
are preferred compared to the scalars. Moreover,  detecting these new fermions 
in the future is more anticipated for a large width of the observed excess.
In conclusion, given this di-photon excess survives
with more data-set, this can not be an isolated {\it surprise}.  Rather, this
must be the pioneering evidence of a BSM mass spectrum while other heavier members
are awaiting to be detected.

\section*{Acknowledgments}

The work of JC is supported by the Department of Science and Technology, 
Government of India under the Grant Agreement number IFA12-PH-34 (INSPIRE Faculty Award). 
The work of AC is supported by the Lancaster-Manchester-Sheffield Consortium for 
Fundamental Physics under STFC Grant No. ST/L000520/1. PG acknowledges
the support from P2IO Excellence Laboratory (LABEX). The work of SM is partially supported by
funding available from the Department of Atomic Energy, Government of India, for the Regional
Centre for Accelerator-based Particle Physics (RECAPP), Harish-Chandra Research Institute.

\bibliographystyle{model1-num-names}
\bibliography{CCGMS-V1.bib}

\end{document}